\documentstyle[12pt]{article}

\newcommand{\sh}{\sinh}
\begin{document}
\immediate\write16{<WARNING: FEYNMAN macros work only with emTeX-dvivers
                    (dviscr.exe, dvihplj.exe, dvidot.exe, etc.) >}
\newdimen\Lengthunit
\newcount\Nhalfperiods
\Lengthunit = 1.5cm
\Nhalfperiods = 9
\catcode`\*=11
\newdimen\L*   \newdimen\d*   \newdimen\d**
\newdimen\dm*  \newdimen\dd*  \newdimen\dt*
\newdimen\a*   \newdimen\b*   \newdimen\c*
\newdimen\a**  \newdimen\b**
\newdimen\xL*  \newdimen\yL*
\newcount\k*   \newcount\l*   \newcount\m*
\newcount\n*   \newcount\dn*  \newcount\r*
\newcount\N*   \newcount\*one \newcount\*two  \*one=1 \*two=2
\newcount\*ths \*ths=1000
\def\GRAPH(hsize=#1)#2{\hbox to #1\Lengthunit{#2\hss}}
\def\Linewidth#1{\special{em:linewidth #1}}
\Linewidth{.4pt}
\def\sm*{\special{em:moveto}}
\def\sl*{\special{em:lineto}}
\newbox\spm*   \newbox\spl*
\setbox\spm*\hbox{\sm*}
\setbox\spl*\hbox{\sl*}
\def\mov#1(#2,#3)#4{\rlap{\L*=#1\Lengthunit\kern#2\L*\raise#3\L*\hbox{#4}}}
\def\smov#1(#2,#3)#4{\rlap{\L*=#1\Lengthunit
\xL*=\xscale\L*\yL*=\yscale\L*\kern#2\xL*\raise#3\yL*\hbox{#4}}}
\def\mov*(#1,#2)#3{\rlap{\kern#1\raise#2\hbox{#3}}}
\def\lin#1(#2,#3){\rlap{\sm*\mov#1(#2,#3){\sl*}}}
\def\arr*(#1,#2,#3){\mov*(#1\dd*,#1\dt*){%
\sm*\mov*(#2\dd*,#2\dt*){\mov*(#3\dt*,-#3\dd*){\sl*}}%
\sm*\mov*(#2\dd*,#2\dt*){\mov*(-#3\dt*,#3\dd*){\sl*}}}}
\def\arrow#1(#2,#3){\rlap{\lin#1(#2,#3)\mov#1(#2,#3){%
\d**=-.012\Lengthunit\dd*=#2\d**\dt*=#3\d**%
\arr*(1,10,4)\arr*(3,8,4)\arr*(4.8,4.2,3)}}}
\def\arrlin#1(#2,#3){\rlap{\L*=#1\Lengthunit\L*=.5\L*%
\lin#1(#2,#3)\mov*(#2\L*,#3\L*){\arrow.1(#2,#3)}}}
\def\dasharrow#1(#2,#3){\rlap{%
{\Lengthunit=0.9\Lengthunit\dashlin#1(#2,#3)\mov#1(#2,#3){\sm*}}%
\mov#1(#2,#3){\sl*\d**=-.012\Lengthunit\dd*=#2\d**\dt*=#3\d**%
\arr*(1,10,4)\arr*(3,8,4)\arr*(4.8,4.2,3)}}}
\def\clap#1{\hbox to 0pt{\hss #1\hss}}
\def\ind(#1,#2)#3{\rlap{%
\d*=.1\Lengthunit\kern#1\d*\raise#2\d*\hbox{\lower2pt\clap{$#3$}}}}
\def\sh*(#1,#2)#3{\rlap{%
\dm*=\the\n*\d**\xL*=\xscale\dm*\yL*=\yscale\dm*
\kern#1\xL*\raise#2\yL*\hbox{#3}}}
\def\calcnum*#1(#2,#3){\a*=1000sp\b*=1000sp\a*=#2\a*\b*=#3\b*%
\ifdim\a*<0pt\a*-\a*\fi\ifdim\b*<0pt\b*-\b*\fi%
\ifdim\a*>\b*\c*=.96\a*\advance\c*.4\b*%
\else\c*=.96\b*\advance\c*.4\a*\fi%
\k*\a*\multiply\k*\k*\l*\b*\multiply\l*\l*%
\m*\k*\advance\m*\l*\n*\c*\r*\n*\multiply\n*\n*%
\dn*\m*\advance\dn*-\n*\divide\dn*2\divide\dn*\r*%
\advance\r*\dn*%
\c*=\the\Nhalfperiods5sp\c*=#1\c*\ifdim\c*<0pt\c*-\c*\fi%
\multiply\c*\r*\N*\c*\divide\N*10000}
\def\dashlin#1(#2,#3){\rlap{\calcnum*#1(#2,#3)%
\d**=#1\Lengthunit\ifdim\d**<0pt\d**-\d**\fi%
\divide\N*2\multiply\N*2\advance\N*1%
\divide\d**\N*\sm*\n*\*one\sh*(#2,#3){\sl*}%
\loop\advance\n*\*one\sh*(#2,#3){\sm*}\advance\n*\*one\sh*(#2,#3){\sl*}%
\ifnum\n*<\N*\repeat}}
\def\dashdotlin#1(#2,#3){\rlap{\calcnum*#1(#2,#3)%
\d**=#1\Lengthunit\ifdim\d**<0pt\d**-\d**\fi%
\divide\N*2\multiply\N*2\advance\N*1\multiply\N*2%
\divide\d**\N*\sm*\n*\*two\sh*(#2,#3){\sl*}\loop%
\advance\n*\*one\sh*(#2,#3){\kern-1.48pt\lower.5pt\hbox{\rm.}}%
\advance\n*\*one\sh*(#2,#3){\sm*}%
\advance\n*\*two\sh*(#2,#3){\sl*}\ifnum\n*<\N*\repeat}}
\def\shl*(#1,#2)#3{\kern#1#3\lower#2#3\hbox{\unhcopy\spl*}}
\def\trianglin#1(#2,#3){\rlap{\toks0={#2}\toks1={#3}\calcnum*#1(#2,#3)%
\dd*=.57\Lengthunit\dd*=#1\dd*\divide\dd*\N*%
\d**=#1\Lengthunit\ifdim\d**<0pt\d**-\d**\fi%
\multiply\N*2\divide\d**\N*\advance\N*-1\sm*\n*\*one\loop%
\shl**{\dd*}\dd*-\dd*\advance\n*2%
\ifnum\n*<\N*\repeat\n*\N*\advance\n*1\shl**{0pt}}}
\def\wavelin#1(#2,#3){\rlap{\toks0={#2}\toks1={#3}\calcnum*#1(#2,#3)%
\dd*=.23\Lengthunit\dd*=#1\dd*\divide\dd*\N*%
\d**=#1\Lengthunit\ifdim\d**<0pt\d**-\d**\fi%
\multiply\N*4\divide\d**\N*\sm*\n*\*one\loop%
\shl**{\dd*}\dt*=1.3\dd*\advance\n*1%
\shl**{\dt*}\advance\n*\*one%
\shl**{\dd*}\advance\n*\*two%
\dd*-\dd*\ifnum\n*<\N*\repeat\n*\N*\shl**{0pt}}}
\def\w*lin(#1,#2){\rlap{\toks0={#1}\toks1={#2}\d**=\Lengthunit\dd*=-.12\d**%
\N*8\divide\d**\N*\sm*\n*\*one\loop%
\shl**{\dd*}\dt*=1.3\dd*\advance\n*\*one%
\shl**{\dt*}\advance\n*\*one%
\shl**{\dd*}\advance\n*\*one%
\shl**{0pt}\dd*-\dd*\advance\n*1\ifnum\n*<\N*\repeat}}
\def\l*arc(#1,#2)[#3][#4]{\rlap{\toks0={#1}\toks1={#2}\d**=\Lengthunit%
\dd*=#3.037\d**\dd*=#4\dd*\dt*=#3.049\d**\dt*=#4\dt*\ifdim\d**>16mm%
\d**=.25\d**\n*\*one\shl**{-\dd*}\n*\*two\shl**{-\dt*}\n*3\relax%
\shl**{-\dd*}\n*4\relax\shl**{0pt}\else\ifdim\d**>5mm%
\d**=.5\d**\n*\*one\shl**{-\dt*}\n*\*two\shl**{0pt}%
\else\n*\*one\shl**{0pt}\fi\fi}}
\def\d*arc(#1,#2)[#3][#4]{\rlap{\toks0={#1}\toks1={#2}\d**=\Lengthunit%
\dd*=#3.037\d**\dd*=#4\dd*\d**=.25\d**\sm*\n*\*one\shl**{-\dd*}%
\n*3\relax\sh*(#1,#2){\xL*=\xscale\dd*\yL*=\yscale\dd*
\kern#2\xL*\lower#1\yL*\hbox{\sm*}}%
\n*4\relax\shl**{0pt}}}
\def\arc#1[#2][#3]{\rlap{\Lengthunit=#1\Lengthunit%
\sm*\l*arc(#2.1914,#3.0381)[#2][#3]%
\smov(#2.1914,#3.0381){\l*arc(#2.1622,#3.1084)[#2][#3]}%
\smov(#2.3536,#3.1465){\l*arc(#2.1084,#3.1622)[#2][#3]}%
\smov(#2.4619,#3.3086){\l*arc(#2.0381,#3.1914)[#2][#3]}}}
\def\dasharc#1[#2][#3]{\rlap{\Lengthunit=#1\Lengthunit%
\d*arc(#2.1914,#3.0381)[#2][#3]%
\smov(#2.1914,#3.0381){\d*arc(#2.1622,#3.1084)[#2][#3]}%
\smov(#2.3536,#3.1465){\d*arc(#2.1084,#3.1622)[#2][#3]}%
\smov(#2.4619,#3.3086){\d*arc(#2.0381,#3.1914)[#2][#3]}}}
\def\wavearc#1[#2][#3]{\rlap{\Lengthunit=#1\Lengthunit%
\w*lin(#2.1914,#3.0381)%
\smov(#2.1914,#3.0381){\w*lin(#2.1622,#3.1084)}%
\smov(#2.3536,#3.1465){\w*lin(#2.1084,#3.1622)}%
\smov(#2.4619,#3.3086){\w*lin(#2.0381,#3.1914)}}}
\def\shl**#1{\c*=\the\n*\d**\d*=#1%
\a*=\the\toks0\c*\b*=\the\toks1\d*\advance\a*-\b*%
\b*=\the\toks1\c*\d*=\the\toks0\d*\advance\b*\d*%
\a*=\xscale\a*\b*=\yscale\b*%
\raise\b*\rlap{\kern\a*\unhcopy\spl*}}
\def\wlin*#1(#2,#3)[#4]{\rlap{\toks0={#2}\toks1={#3}%
\c*=#1\l*\c*\c*=.01\Lengthunit\m*\c*\divide\l*\m*%
\c*=\the\Nhalfperiods5sp\multiply\c*\l*\N*\c*\divide\N*\*ths%
\divide\N*2\multiply\N*2\advance\N*1%
\dd*=.002\Lengthunit\dd*=#4\dd*\multiply\dd*\l*\divide\dd*\N*%
\d**=#1\multiply\N*4\divide\d**\N*\sm*\n*\*one\loop%
\shl**{\dd*}\dt*=1.3\dd*\advance\n*\*one%
\shl**{\dt*}\advance\n*\*one%
\shl**{\dd*}\advance\n*\*two%
\dd*-\dd*\ifnum\n*<\N*\repeat\n*\N*\shl**{0pt}}}
\def\wavebox#1{\setbox0\hbox{#1}%
\a*=\wd0\advance\a*14pt\b*=\ht0\advance\b*\dp0\advance\b*14pt%
\hbox{\kern9pt%
\mov*(0pt,\ht0){\mov*(-7pt,7pt){\wlin*\a*(1,0)[+]\wlin*\b*(0,-1)[-]}}%
\mov*(\wd0,-\dp0){\mov*(7pt,-7pt){\wlin*\a*(-1,0)[+]\wlin*\b*(0,1)[-]}}%
\box0\kern9pt}}
\def\rectangle#1(#2,#3){%
\lin#1(#2,0)\lin#1(0,#3)\mov#1(0,#3){\lin#1(#2,0)}\mov#1(#2,0){\lin#1(0,#3)}}
\def\dashrectangle#1(#2,#3){\dashlin#1(#2,0)\dashlin#1(0,#3)%
\mov#1(0,#3){\dashlin#1(#2,0)}\mov#1(#2,0){\dashlin#1(0,#3)}}
\def\waverectangle#1(#2,#3){\L*=#1\Lengthunit\a*=#2\L*\b*=#3\L*%
\ifdim\a*<0pt\a*-\a*\def\x*{-1}\else\def\x*{1}\fi%
\ifdim\b*<0pt\b*-\b*\def\y*{-1}\else\def\y*{1}\fi%
\wlin*\a*(\x*,0)[-]\wlin*\b*(0,\y*)[+]%
\mov#1(0,#3){\wlin*\a*(\x*,0)[+]}\mov#1(#2,0){\wlin*\b*(0,\y*)[-]}}
\def\calcparab*{%
\ifnum\n*>\m*\k*\N*\advance\k*-\n*\else\k*\n*\fi%
\a*=\the\k* sp\a*=10\a*\b*\dm*\advance\b*-\a*\k*\b*%
\a*=\the\*ths\b*\divide\a*\l*\multiply\a*\k*%
\divide\a*\l*\k*\*ths\r*\a*\advance\k*-\r*%
\dt*=\the\k*\L*}
\def\arcto#1(#2,#3)[#4]{\rlap{\toks0={#2}\toks1={#3}\calcnum*#1(#2,#3)%
\dm*=135sp\dm*=#1\dm*\d**=#1\Lengthunit\ifdim\dm*<0pt\dm*-\dm*\fi%
\multiply\dm*\r*\a*=.3\dm*\a*=#4\a*\ifdim\a*<0pt\a*-\a*\fi%
\advance\dm*\a*\N*\dm*\divide\N*10000%
\divide\N*2\multiply\N*2\advance\N*1%
\L*=-.25\d**\L*=#4\L*\divide\d**\N*\divide\L*\*ths%
\m*\N*\divide\m*2\dm*=\the\m*5sp\l*\dm*%
\sm*\n*\*one\loop\calcparab*\shl**{-\dt*}%
\advance\n*1\ifnum\n*<\N*\repeat}}
\def\arrarcto#1(#2,#3)[#4]{\L*=#1\Lengthunit\L*=.54\L*%
\arcto#1(#2,#3)[#4]\mov*(#2\L*,#3\L*){\d*=.457\L*\d*=#4\d*\d**-\d*%
\mov*(#3\d**,#2\d*){\arrow.02(#2,#3)}}}
\def\dasharcto#1(#2,#3)[#4]{\rlap{\toks0={#2}\toks1={#3}\calcnum*#1(#2,#3)%
\dm*=\the\N*5sp\a*=.3\dm*\a*=#4\a*\ifdim\a*<0pt\a*-\a*\fi%
\advance\dm*\a*\N*\dm*%
\divide\N*20\multiply\N*2\advance\N*1\d**=#1\Lengthunit%
\L*=-.25\d**\L*=#4\L*\divide\d**\N*\divide\L*\*ths%
\m*\N*\divide\m*2\dm*=\the\m*5sp\l*\dm*%
\sm*\n*\*one\loop%
\calcparab*\shl**{-\dt*}\advance\n*1%
\ifnum\n*>\N*\else\calcparab*%
\sh*(#2,#3){\kern#3\dt*\lower#2\dt*\hbox{\sm*}}\fi%
\advance\n*1\ifnum\n*<\N*\repeat}}
\def\*shl*#1{%
\c*=\the\n*\d**\advance\c*#1\a**\d*\dt*\advance\d*#1\b**%
\a*=\the\toks0\c*\b*=\the\toks1\d*\advance\a*-\b*%
\b*=\the\toks1\c*\d*=\the\toks0\d*\advance\b*\d*%
\raise\b*\rlap{\kern\a*\unhcopy\spl*}}
\def\calcnormal*#1{%
\b**=10000sp\a**\b**\k*\n*\advance\k*-\m*%
\multiply\a**\k*\divide\a**\m*\a**=#1\a**\ifdim\a**<0pt\a**-\a**\fi%
\ifdim\a**>\b**\d*=.96\a**\advance\d*.4\b**%
\else\d*=.96\b**\advance\d*.4\a**\fi%
\d*=.01\d*\r*\d*\divide\a**\r*\divide\b**\r*%
\ifnum\k*<0\a**-\a**\fi\d*=#1\d*\ifdim\d*<0pt\b**-\b**\fi%
\k*\a**\a**=\the\k*\dd*\k*\b**\b**=\the\k*\dd*}
\def\wavearcto#1(#2,#3)[#4]{\rlap{\toks0={#2}\toks1={#3}\calcnum*#1(#2,#3)%
\c*=\the\N*5sp\a*=.4\c*\a*=#4\a*\ifdim\a*<0pt\a*-\a*\fi%
\advance\c*\a*\N*\c*\divide\N*20\multiply\N*2\advance\N*-1\multiply\N*4%
\d**=#1\Lengthunit\dd*=.012\d**\ifdim\d**<0pt\d**-\d**\fi\L*=.25\d**%
\divide\d**\N*\divide\dd*\N*\L*=#4\L*\divide\L*\*ths%
\m*\N*\divide\m*2\dm*=\the\m*0sp\l*\dm*%
\sm*\n*\*one\loop\calcnormal*{#4}\calcparab*%
\*shl*{1}\advance\n*\*one\calcparab*%
\*shl*{1.3}\advance\n*\*one\calcparab*%
\*shl*{1}\advance\n*2%
\dd*-\dd*\ifnum\n*<\N*\repeat\n*\N*\shl**{0pt}}}
\def\triangarcto#1(#2,#3)[#4]{\rlap{\toks0={#2}\toks1={#3}\calcnum*#1(#2,#3)%
\c*=\the\N*5sp\a*=.4\c*\a*=#4\a*\ifdim\a*<0pt\a*-\a*\fi%
\advance\c*\a*\N*\c*\divide\N*20\multiply\N*2\advance\N*-1\multiply\N*2%
\d**=#1\Lengthunit\dd*=.012\d**\ifdim\d**<0pt\d**-\d**\fi\L*=.25\d**%
\divide\d**\N*\divide\dd*\N*\L*=#4\L*\divide\L*\*ths%
\m*\N*\divide\m*2\dm*=\the\m*0sp\l*\dm*%
\sm*\n*\*one\loop\calcnormal*{#4}\calcparab*%
\*shl*{1}\advance\n*2%
\dd*-\dd*\ifnum\n*<\N*\repeat\n*\N*\shl**{0pt}}}
\def\hr*#1{\clap{\xL*=\xscale\Lengthunit\vrule width#1\xL* height.1pt}}
\def\shade#1[#2]{\rlap{\Lengthunit=#1\Lengthunit%
\smov(0,#2.05){\hr*{.994}}\smov(0,#2.1){\hr*{.980}}%
\smov(0,#2.15){\hr*{.953}}\smov(0,#2.2){\hr*{.916}}%
\smov(0,#2.25){\hr*{.867}}\smov(0,#2.3){\hr*{.798}}%
\smov(0,#2.35){\hr*{.715}}\smov(0,#2.4){\hr*{.603}}%
\smov(0,#2.45){\hr*{.435}}}}
\def\dshade#1[#2]{\rlap{%
\Lengthunit=#1\Lengthunit\if#2-\def\t*{+}\else\def\t*{-}\fi%
\smov(0,\t*.025){%
\smov(0,#2.05){\hr*{.995}}\smov(0,#2.1){\hr*{.988}}%
\smov(0,#2.15){\hr*{.969}}\smov(0,#2.2){\hr*{.937}}%
\smov(0,#2.25){\hr*{.893}}\smov(0,#2.3){\hr*{.836}}%
\smov(0,#2.35){\hr*{.760}}\smov(0,#2.4){\hr*{.662}}%
\smov(0,#2.45){\hr*{.531}}\smov(0,#2.5){\hr*{.320}}}}}
\def\vdot{\rlap{\kern-1.9pt\lower1.8pt\hbox{$\scriptstyle\bullet$}}}
\def\vtimes{\rlap{\kern-3pt\lower1.8pt\hbox{$\scriptstyle\times$}}}
\def\vDot{\rlap{\kern-2.3pt\lower2.7pt\hbox{$\bullet$}}}
\def\vTimes{\rlap{\kern-3.6pt\lower2.4pt\hbox{$\times$}}}
\catcode`\*=12
\newcount\CatcodeOfAtSign
\CatcodeOfAtSign=\the\catcode`\@
\catcode`\@=11
\newcount\n@ast
\def\n@ast@#1{\n@ast0\relax\get@ast@#1\end}
\def\get@ast@#1{\ifx#1\end\let\next\relax\else%
\ifx#1*\advance\n@ast1\fi\let\next\get@ast@\fi\next}
\newif\if@up \newif\if@dwn
\def\up@down@#1{\@upfalse\@dwnfalse%
\if#1u\@uptrue\fi\if#1U\@uptrue\fi\if#1+\@uptrue\fi%
\if#1d\@dwntrue\fi\if#1D\@dwntrue\fi\if#1-\@dwntrue\fi}
\def\halfcirc#1(#2)[#3]{{\Lengthunit=#2\Lengthunit\up@down@{#3}%
\if@up\smov(0,.5){\arc[-][-]\arc[+][-]}\fi%
\if@dwn\smov(0,-.5){\arc[-][+]\arc[+][+]}\fi%
\def\lft{\smov(0,.5){\arc[-][-]}\smov(0,-.5){\arc[-][+]}}%
\def\rght{\smov(0,.5){\arc[+][-]}\smov(0,-.5){\arc[+][+]}}%
\if#3l\lft\fi\if#3L\lft\fi\if#3r\rght\fi\if#3R\rght\fi%
\n@ast@{#1}%
\ifnum\n@ast>0\if@up\shade[+]\fi\if@dwn\shade[-]\fi\fi%
\ifnum\n@ast>1\if@up\dshade[+]\fi\if@dwn\dshade[-]\fi\fi}}
\def\halfdashcirc(#1)[#2]{{\Lengthunit=#1\Lengthunit\up@down@{#2}%
\if@up\smov(0,.5){\dasharc[-][-]\dasharc[+][-]}\fi%
\if@dwn\smov(0,-.5){\dasharc[-][+]\dasharc[+][+]}\fi%
\def\lft{\smov(0,.5){\dasharc[-][-]}\smov(0,-.5){\dasharc[-][+]}}%
\def\rght{\smov(0,.5){\dasharc[+][-]}\smov(0,-.5){\dasharc[+][+]}}%
\if#2l\lft\fi\if#2L\lft\fi\if#2r\rght\fi\if#2R\rght\fi}}
\def\halfwavecirc(#1)[#2]{{\Lengthunit=#1\Lengthunit\up@down@{#2}%
\if@up\smov(0,.5){\wavearc[-][-]\wavearc[+][-]}\fi%
\if@dwn\smov(0,-.5){\wavearc[-][+]\wavearc[+][+]}\fi%
\def\lft{\smov(0,.5){\wavearc[-][-]}\smov(0,-.5){\wavearc[-][+]}}%
\def\rght{\smov(0,.5){\wavearc[+][-]}\smov(0,-.5){\wavearc[+][+]}}%
\if#2l\lft\fi\if#2L\lft\fi\if#2r\rght\fi\if#2R\rght\fi}}
\def\Circle#1(#2){\halfcirc#1(#2)[u]\halfcirc#1(#2)[d]\n@ast@{#1}%
\ifnum\n@ast>0\clap{%
\dimen0=\xscale\Lengthunit\vrule width#2\dimen0 height.1pt}\fi}
\def\wavecirc(#1){\halfwavecirc(#1)[u]\halfwavecirc(#1)[d]}
\def\dashcirc(#1){\halfdashcirc(#1)[u]\halfdashcirc(#1)[d]}
%
\def\xscale{1}
\def\yscale{1}
\def\Ellipse#1(#2)[#3,#4]{\def\xscale{#3}\def\yscale{#4}%
\Circle#1(#2)\def\xscale{1}\def\yscale{1}}
\def\dashEllipse(#1)[#2,#3]{\def\xscale{#2}\def\yscale{#3}%
\dashcirc(#1)\def\xscale{1}\def\yscale{1}}
\def\waveEllipse(#1)[#2,#3]{\def\xscale{#2}\def\yscale{#3}%
\wavecirc(#1)\def\xscale{1}\def\yscale{1}}
\def\halfEllipse#1(#2)[#3][#4,#5]{\def\xscale{#4}\def\yscale{#5}%
\halfcirc#1(#2)[#3]\def\xscale{1}\def\yscale{1}}
\def\halfdashEllipse(#1)[#2][#3,#4]{\def\xscale{#3}\def\yscale{#4}%
\halfdashcirc(#1)[#2]\def\xscale{1}\def\yscale{1}}
\def\halfwaveEllipse(#1)[#2][#3,#4]{\def\xscale{#3}\def\yscale{#4}%
\halfwavecirc(#1)[#2]\def\xscale{1}\def\yscale{1}}
\catcode`\@=\the\CatcodeOfAtSign


\hfill hep-th/9905062

\vspace*{1cm}

\begin{center}
{\LARGE
Holomorphic effective potential in general chiral superfield model}
\end{center}

\centerline{\large I.L. Buchbinder
and A.Yu. Petrov}
\begin{center}
{\small\it Department
of Theoretical Physics, Tomsk State Pedagogical University}\\
 {\small\it 634041
Tomsk, Russia}

\end{center}

\vspace*{.5cm}

\begin{abstract}
We study a holomorphic effective potential $W_{eff}(\Phi)$ in chiral superfield
model defined in terms of
arbitrary k\"{a}hlerian potential
$K(\bar{\Phi},\Phi)$ and arbitrary chiral potential $W(\Phi)$.
Such a model naturally arises as an ingredient of low-energy limit of
superstring theory and it is called here the general chiral superfield model.
Generic procedure for calculating the chiral loop corrections to
effective action is developed. We find lower two-loop chiral correction
in the form
$W_{eff}^{(2)}(\Phi)=\frac{6}{{(4\pi)}^4}\zeta(3)\bar{W}^{'''2}(0)
{\big(\frac{W^{''}(\Phi)}{K^2_{\Phi\bar{\Phi}}(0,\Phi)}
\big)}^3$ where
$K_{\Phi\bar{\Phi}}(0,\Phi)=\frac{\partial^2 K(\bar{\Phi},\Phi)}
{\partial\Phi\partial\bar{\Phi}}|_{\bar{\Phi}=0}$ and $\zeta (x)$ be\\
Riemannian dzeta-function. This correction is finite at any
$K(\bar{\Phi},\Phi)$, $W(\Phi)$.
\end{abstract}

According to the superstring theory the low-energy elementary particle
models
contain as ingredient the multiplets of chiral and antichiral superfields action
of which is given in terms of k\"{a}hlerian effective potential
$K(\bar{\Phi},\Phi)$ and chiral $W(\Phi)$ and antichiral $\bar{W}(\bar{\Phi})$
potentials. These potentials are found in explicit and closed form
within string perturbation theory (see f.e. \cite{GSW}).
Phenomenological aspects of such models have been studied in recent papers
\cite{Cv}. In quantum theory one can expect an appearance of quantum corrections
to the potentials $K(\bar{\Phi},\Phi)$ and $W(\Phi)$. As a result we face a
problem of calculating effective action in models with arbitrary functions
$K(\bar{\Phi},\Phi)$ and $W(\Phi)$.

Manifestly supersymmetric technique for calculations of effective action in the
theories with $N=1$ chiral superfields was formulated in recent refs.
\cite{Buch1}--\cite{Buch4}. The new approaches have been developed in refs.
\cite{GR}.

The remarkable features of the massless theories with $N=1$ chiral
superfields are the possibilities of obtaining the chiral quantum
corrections.  A few years ago West \cite{West2} pointed out that finite
two-loop chiral contribution to effective action really arises in
massless Wess-Zumino model (see also \cite{West3,Buch3}).

In this paper we solve the general problem of calculating leading quantum
correction to chiral potential in massless theory with
arbitrary potentials $K(\bar{\Phi},\Phi)$ and $W(\Phi)$, $\bar{W}(\bar{\Phi})$.
Since all such corrections depend only on $\Phi$ but not on $\bar{\Phi}$ they
form an effective potential which can be called chiral or holomorphic effective
potential unlike of k\"{a}hlerian effective potential depending both on
$\Phi$ and $\bar{\Phi}$.
The remarkable result we obtain in this paper
is that despite the theory under consideration is non-renormalizable at
arbitrary $K(\bar{\Phi},\Phi)$, $W(\Phi)$,
the lower (two-loop) holomorphic correction to effective action is always
finite.


We consider $N=1$ supersymmetric field theory with action
\begin{equation}
\label{act}
S[\bar{\Phi},\Phi]=\int d^8 z K(\bar{\Phi},\Phi)+
(\int d^6 z W(\Phi)+h.c.)
\end{equation}
where $\Phi(z)$ and $\bar{\Phi}(z)$ are chiral and antichiral superfields
respectively. As well known, the real function $K(\bar{\Phi},\Phi)$
is called k\"{a}hlerian potential and holomorphic function $W(\Phi)$ is
called chiral potential \cite{BK0}. The partial cases of the theory (1) are
Wess-Zumino model with $K(\bar{\Phi},\Phi)=\Phi\bar{\Phi}$, $W(\Phi)\sim\Phi^3$
and $N=1$ supersymmetric four-dimensional sigma-model with $W(\Phi)=0$.
The action (1) is a most general one constructed from chiral and antichiral
superfields which does not contain the higher derivatives at component level.
Therefore it is natural to call the theory (1) a general chiral superfield
model. In this letter we are going to calculate the lowest quantum corrections
to chiral potential $W(\Phi)$ in the model under consideration.

Let $\Gamma [\Phi, \bar{\Phi}]$ be effective action in the model (1).
Within a momentum expansion the effective action can be presented as a series
in supercovariant derivatives
$D_A=(\partial_a,D_{\alpha},\bar{D}_{\dot{\alpha}})$ in the form
\begin{equation}
 \Gamma[\bar{\Phi},\Phi] = \int d^8z {\cal L}_{eff}
(\Phi,D_A\Phi,D_A D_B\Phi;\bar{\Phi},D_A\bar{\Phi},D_A D_B\bar{\Phi})
+ (\int d^6z {\cal L}^{(c)}_{eff}(\Phi) + h.c.) +\ldots
\end{equation}
Here
${\cal L}_{eff}$ is called general effective lagrangian,
${\cal L}^{(c)}_{eff}$ is called chiral effective lagrangian. Both these
lagrangians are the series in supercovariant derivatives
of superfields and can be written as follows
\begin{eqnarray}
\label{ep}
{\cal L}_{eff}&=&K_{eff}(\bar{\Phi},\Phi)+\ldots
=K(\bar{\Phi},\Phi)+\sum_{n=1}^{\infty}K^{(n)}_{eff}(\bar{\Phi},\Phi)\nonumber\\
{\cal L}^{(c)}_{eff}&=&W_{eff}(\Phi)+\ldots=
W(\Phi)+\sum_{n=1}^{\infty}W^{(n)}_{eff}(\Phi)+\ldots
\end{eqnarray}
Here dots mean terms depending on covariant derivatives
of superfields
$\Phi,\bar{\Phi}$. Here $K_{eff}(\bar{\Phi},\Phi)$ is called
kahlerian effective potential, $W_{eff}(\Phi)$ is called chiral effective
potential or holomorphic effective potential,
$K^{(n)}_{eff}$ is a $n$-th correction to kahlerian potential
and $W^{(n)}_{eff}$ is a $n$-th correcton to chiral or holomorphic
potential $W$.

To consider the effective lagrangians ${\cal L}_{eff}$ and
${\cal L}^{(c)}_{eff}$ we consider the path integral representation of
the effective action \cite{BK0,BO}
\begin{eqnarray}
\label{Green1}
 \exp(\frac{i}{\hbar}\Gamma[\bar{\Phi},\Phi]) &=&
 \int {\cal D} \phi {\cal D} \bar{\phi}
 \exp\big(
\frac{i}{\hbar}
S[\bar{\Phi}+\sqrt{\hbar}\bar{\phi},\Phi+\sqrt{\hbar}\phi]
-\nonumber\\&-&
(\int d^6 z
\frac{\delta\Gamma[\bar{\Phi},\Phi]}{\delta\Phi(z)}\phi(z)+h.c. )
\big)
\end{eqnarray}
Here $\Phi,\bar{\Phi}$ are the background superfields and $\phi,\bar{\phi}$
are the quantum ones.
The effective action can be written as
$\Gamma[\bar{\Phi},\Phi]=S[\bar{\Phi},\Phi]+\tilde{\Gamma}[\bar{\Phi},\Phi]$,
where $\tilde{\Gamma}[\bar{\Phi},\Phi]$ is a quantum correction.
Eq. (4) allows to obtain $\tilde{\Gamma}[\bar{\Phi},\Phi]$ in form of
loop expansion \begin{equation} \label{Gamma}
 \tilde{\Gamma}[\bar{\Phi},\Phi] = \sum_{n=1}^{\infty}\hbar^n
\Gamma^{(n)} [\bar{\Phi},\Phi]
\end{equation}
and hence, to get loop expansion for the effective lagrangians
${\cal L}_{eff}$ and ${\cal L}^{(c)}_{eff}$

To find loop corrections $\Gamma^{(n)} [\bar{\Phi},\Phi]$ in explicit
 form we expand right-hand side of eq. (4) in power series in quantum
superfields $\phi$, $\bar{\phi}$. As usual, the quadratic part of
expansion of
$\frac{1}{\hbar}S[\bar{\Phi}+\sqrt{\hbar}\bar{\phi},\Phi+\sqrt{\hbar}\phi] $
\begin{eqnarray}
\label{qua}
S_2&=&\frac{1}{2}\int d^8 z \left(\begin{array}{cc}\phi&\bar{\phi}
\end{array}\right)
\left(\begin{array}{cc}
K_{\Phi\Phi}&K_{\Phi\bar{\Phi}}\\
K_{\Phi\bar{\Phi}}&K_{\bar{\Phi}\bar{\Phi}}
\end{array}\right)
\left(\begin{array}{c}\phi\\
\bar{\phi}
\end{array}
\right)+[\int d^6 z \frac{1}{2}W^{''}\phi^2+h.c.]
\end{eqnarray}
defines the propagators and the higher terms of expansion define the vertices.
Here
$K_{\Phi\bar{\Phi}}=\frac{\partial^2 K(\bar{\Phi},\Phi)}
{\partial\Phi\partial\bar{\Phi}}$,
$K_{\Phi\Phi}=\frac{\partial^2 K(\bar{\Phi},\Phi)}
{\partial\Phi^2}$ etc,  $W^{''}=\frac{d^2 W}{d\Phi^2}$.

The theory under consideration is
characterized by matrix superpropagator
$G$
\begin{equation}
\label{gre}
 G(z_1,z_2) =
 \left(
 \begin{array}{ll}
  G_{++}(z_1,z_2) & G_{+-}(z_1,z_2)\\
  G_{-+}(z_1,z_2) & G_{--}(z_1,z_2)
 \end{array}
 \right)
\end{equation}
Equation of motion for this propagator is obtained from action $S_2$
(\ref{qua}) in the form
\begin{equation}
\label{def}
 \left(
 \begin{array}{cc}
  W''-\frac{1}{4}(\bar{D}^2 K_{\Phi\Phi}) &-\frac{1}{4}\vec{\bar{D}^2} K_{\Phi\bar{\Phi}}\\
-\frac{1}{4}\vec{D^2}  K_{\Phi\bar{\Phi}} &
\bar{W''}-\frac{1}{4}(D^2 K_{\bar{\Phi}\bar{\Phi}})
 \end{array}
 \right)
 \left(
 \begin{array}{ll}
  G_{++} & G_{+-}\\
  G_{-+} & G_{--}
 \end{array}
 \right)
 =
-\left(
 \begin{array}{ll}
  \delta_+ & 0\\
  0 & \delta_-
 \end{array}
 \right)
\end{equation}
Here the arrows mean that the corresponding operators act on all function at the
right side, and
$\delta_+=-\frac{1}{4}\bar{D}^2\delta^8(z_1-z_2)$,
$\delta_-=-\frac{1}{4}D^2\delta^8(z_1-z_2)$ are chiral and antichiral delta
functions respectively. The signs $+$ and $-$ at the matrix elements of
$G(z_1,z_2)$ denote that this matrix element is chiral
$(+)$ or antichiral $(-)$ in corresponding argument.
It is evident the superpropagator (\ref{gre})
depends on background superfields $\Phi$, $\bar{\Phi}$.
Since we are going
to obtain the quantum corrections to holomorphic potential we can put
$\bar{\Phi}=0$ in eq. (\ref{def}).


Let us consider a general procedure of calculating the holomorphic
effective potential $W_{eff}(\Phi)$. The mechanism of arising the chiral
corrections to effective action looks as follows. According to
non-renormalization theorem (see f.e. \cite{BK0}) all loop corrections to
effective action can be expressed in form
of integral over full superspace.
However, it was pointed by West \cite{West2}
that non-renormalization theorem does not forbid an existence of the finite
chiral corrections to the effective action. The matter is the theories with chiral
superfields admit the loop corrections of the form
\begin{equation}
\label{chr}
\int d^8 z f(\Phi)\big(-\frac{D^2}{4\Box}\big)g(\Phi)=\int d^6 z f(\Phi)g(\Phi)
\end{equation}
where $f(\Phi),g(\Phi)$ are some functions of chiral superfield $\Phi$.
We note that eq. (\ref{chr}) shows the
superfield $\Phi$ is not a constant and if we are going to use eq. (\ref{chr})
we should keep all terms with covariant derivatives in eq. (\ref{def}).

It is easy to understand why the chiral corrections can arise only for massless
theories. Te factor
$\Box^{-1}$ in (\ref{chr}) can originate only from massless propagators.
In massive cases the propagators looks like
${(\Box-m^2)}^{-1}$ and the left-hand side of eq.(\ref{chr}) will have the
following form
\begin{equation}
\label{chm}
\int d^8 z f(\Phi)\big(-\frac{D^2}{4(\Box-m^2)}\big)g(\Phi)
\end{equation}
instead of (\ref{chr}).
After transformation to integral over chiral
subspace and using Fourier transformation one gets
\begin{equation}
\label{chm1}
\int\frac{d^4 p}{{(2\pi)}^4}f(\Phi)\big(\frac{p^2}{p^2+m^2}\big)g(\Phi)
\end{equation}
Since we are going to obtain just potential we should consider
the superfields slowly varying in space-time. It means we take the limit
$p^2\to 0$. It is evident that expression (\ref{chm1}) vanishes at
$m^2\neq 0$, $p^2\to 0$.
Therefore non-trivial chiral corrections
can arise only in massless theory.

To find chiral corrections to effective action we put
$\bar{\Phi}=0$ in eqs. (4,5,7).
{\bf Therefore here and further all derivatives of $K$, $W$ and $\bar{W}$ will
be taken at $\bar{\Phi}=0$.} Under this condition
the action of quantum superfields $\phi,\bar{\phi}$ in external superfield
$\Phi$ looks like
\begin{eqnarray}
\label{qch}
S[\bar{\phi},\phi,\Phi]&=&\frac{1}{2}\int d^8 z
\left(\begin{array}{cc}\phi&\bar{\phi} \end{array}\right)
\left(\begin{array}{cc}
K_{\Phi\Phi}&K_{\Phi\bar{\Phi}}\\
K_{\Phi\bar{\Phi}}&K_{\bar{\Phi}\bar{\Phi}}
\end{array}\right)
\left(\begin{array}{c}\phi\\
\bar{\phi}
\end{array}
\right)+\int d^6 z \frac{1}{2}W^{''}\phi^2+\ldots
\end{eqnarray}
The dots here denote the terms of third, fourth and higher orders in quantum
superfields.
We  call the theory massless if $W^{''}|_{\Phi=0}=0$.
Further we consider only massless theory.

To calculate the corrections to $W(\Phi)$ we use supergraph technique
(see f.e. \cite{BK0}).
For this purpose one splits the action (\ref{qch}) into sum of free
part and vertices of interaction.
As a free part we take the action
$$
S_0=\int d^8 z \phi\bar{\phi}
$$
The corresponding superpropagator is
\begin{equation}
\label{pr1}
G(z_1,z_2) = -\frac{D^2_1\bar{D}^2_2}{16\Box}\delta^8(z_1-z_2)
\end{equation}
And the term $S[\bar{\phi},\phi,\Phi]-S_0$
will be treated as vertices where $S[\bar{\phi},\phi,\Phi]$
is given by eq. (10). Our purpose is to find first leading contribution to
$W_{eff}(\Phi)$. As we will show, chiral loop contributions are began with two
loops. Therefore we keep in eq. (10)  only the terms of second, third and fourth
orders in quantum fields.

According to (improved) supergraph technique \cite{Grisaru2} (see also
\cite{BK0}) it is convenient to associate all possible $D$-factors in
diagrams with propagators but not with vertices and transform all
contributions in the form with a single integral over Grassmannian
coordinates. Since the action (10) contains the pure chiral vertices we
should use one of factors $\bar{D}^2$ or $D^2$ in the propagators to
transform the integrals over chiral subspace into the integrals over
full superspace.  As a result one of propagators associated with each
such a vertex contains one factor $\bar{D}^2$ or $D^2$ less (see
\cite{BK0} for details).

The possible vertices contributing to one- and two-loop
corrections have the structure
\begin{eqnarray}
\label{vert}
& &K_{\bar{\Phi}\bar{\Phi}}\bar{\phi}^2,
\ (K_{\Phi\bar{\Phi}}-1)\phi\bar{\phi},\ \frac{1}{2}W^{''}\phi^2,
W_{\Phi\Phi\Phi}\phi^3,\ W_{\Phi\Phi\Phi\Phi}\phi^4,
K_{\Phi\Phi\bar{\Phi}}\phi^2\bar{\phi},\
\nonumber\\
& &
\ K_{\Phi\Phi\bar{\Phi}\bar{\Phi}}\phi^2\bar{\phi}^2,\
K_{\Phi\Phi\Phi\bar{\Phi}}\phi^3\bar{\phi}
\end{eqnarray}
There should be also the vertices conjugate to ones (12).

To clarify which vertices actually contribute to chiral quantum corrections
we carry out a dimensional analysis of the supergraphs in one- and two-loop
approximations. To be more precise we find full mass dimension of the
supergraphs depending on number of vertices, propagators
and loop integrations in the diagram.

Each propagator of the form (\ref{pr1}) contributes 0 since dimension
of $D^{\alpha}$, $\bar{D}_{\dot{\alpha}}$ is equal to $1/2$, and that one of
$\partial^m$ is equal to 1.

Each loop includes integration over momenta with contribution to
dimension equal to 4.
Then, each contraction of a loop into a point in $\theta$-space by the
rule $\delta_{12} D^2_1\bar{D}^2_1\delta_{12}=16\delta_{12}$ requires four
$D,\bar{D}$-factors. Therefore total number of $D^{\alpha}$-,
$\bar{D}_{\dot{\alpha}}$-factors contributing to the dimension
is reduced by 4, and hence each loop contributes 2.

Each vertex proportional to $W^{''}$ ($\bar{W}^{''}$) formally corresponds to
two $\bar{D}^2$- ($D^2$-) factors but one of them is used to transform
the contribution of the vertex to the form of an integral over $d^8 z$.
Therefore each such a vertex reduces mass dimension by 1. One can show
analogously that each vertex proportional to higher derivatives of
$W$, $\bar{W}$ reduces mass dimension by 1.

One of factors $D^2$ associated with propagators
is transformed to d'Alembertian operator when the
expression (\ref{chr}) is converted to the form of an integral over chiral
subspace. Hence total dimension of the supergraph is increased by 1.

Therefore contribution of each diagram has the dimension equal to
$2L+1-n_{W^{''}}-n_{V_c}$ where $L$ is a number of loops, $n_{W^{''}}$
is a number of vertices $W^{''}$, $n_{V_c}$
is a number of vertices proportional to third and higher derivatives of $W$,
$\bar{W}$.
By definition effective potential is an effective lagrangian in limit
$p^2\to 0$. Therefore non-trivial correction to holomorphic effective
potential can arise only at \begin{equation} \label{ind}
2L+1-n_{W^{''}}-n_{V_c}=0
\end{equation}
Otherwise the contribution from diagram can either vanish or be singular in
infrared limit. We note that vertices proportional to derivatives from
$K(\bar{\Phi},\Phi)$ do not contribute to dimension.

After each pair
$\bar{D}^2 D^2$ is transformed to square of internal momentum
by the rule $\bar{D}^2 D^2=-16 k^2$ one
factor $D^2$ should rest to be converted to square of external momentum
after transformation of the expression (\ref{chr}) to the form of integral over
chiral subspace. Therefore diagram contributes to holomorphic effective
potential if and only if the number of $D^2$-factors is more by one
than the number of $\bar{D}^2$'s after transformation of all vertices
to the form of integrals over full superspace.

It is easy to see that one of two factors $\bar{D}^2$ associated with the vertex
proportional to $K_{\Phi\Phi}$ can be transported to external line only. Since
all derivatives of $K$ and $W$ are considered at
$\bar{\Phi}=0$ the acting of $\bar{D}^2$ to the external line $K_{\Phi\Phi}$
leads to zero.
Hence vertices proportional to $K_{\Phi\Phi}$ do not contribute to
holomorphic effective potential.

Now let us consider different one-loop and two-loop diagrams. In one-loop
approximation $n_{W^{''}}+n_{V_c}=3$ because of (\ref{ind}). However, for
one-loop supergraphs $V_c=0$, therefore $n_{W^{''}}=3$. Since each of external
vertices $W^{''}$ corresponds to $\bar{D}^2$,
and number of $D^2$-factors must be more by one than number of
$\bar{D}^2$-factors, possible diagram should contain
two vertices proportional to $K_{\bar{\Phi}\bar{\Phi}}$. However,
straightforward constructing shows that the
one-loop diagram of such a type should contain a line proportional to
$\frac{\bar{D}^2_1\bar{D}^2_2}{16}\delta_{12}=0$. Therefore contribution of
this supergraph is equal to zero, and one-loop correction to
holomorphic effective potential is absent at all, $W^{(1)}(\Phi)=0$.

In two-loop approximation one gets the condition $n_{W^{''}}+n_{V_c}=5$
because of (\ref{ind}). Since the number of purely chiral (antichiral)
vertices independent of $W^{''}$ in two-loop supergraphs can be equal
to 0, 1 or 2, number of external vertices $W^{''}$ takes values from 3
to 5.

It is well known that there are two types of two-loop one-particle-irreducible
diagrams (see Figs. 1a, 1b).

\vspace*{2mm}

\hspace{2cm}
\Lengthunit=1.2cm
\GRAPH(hsize=3){\ind(10,-15){Fig.1a}
\mov(.5,0){\Circle(2)\mov(2,0){\Circle(2)}}
\mov(4,0){\GRAPH(hsize=3){
\ind(15,-15){Fig.1b}
\mov(2,0){\Circle(2)\mov(-1,0){\lin(2,0)}}
}
}
}

\vspace*{2mm}

\noindent However, to consider contributions from diagrams of these types
we must take into account that they can include all vertices described in
(\ref{vert}).
It was mentioned already that each vertex proportional to
$W^{''}$ corresponds to one factor $\bar{D}^2$, and each chiral (antichiral)
vertex proportional to $\phi^n$ ($\bar{\phi}^n$) corresponds to
$n-1$ factors $\bar{D}^2$ ($D^2$)
\footnote{We note that each vertex of such types corresponds to an integral
over $d^6 z$ ($d^6\bar{z}$). At the same time, one factor $\bar{D}^2$ ($D^2$)
is associated with each field $\phi$ ($\bar{\phi}$). Hence vertex $\phi^n$
($\bar{\phi}^n$) corresponds to $n$ factors $\bar{D}^2$ ($D^2$) but one of them
is used to transform the expression to the form of an integral over $d^8 z$.}.
 Each vertex proportional to
$\phi^m\bar{\phi}^n$ corresponds to $m$ factors $\bar{D}^2$ and $n$
factors $D^2$.

We note that non-trivial contribution to holomorphic
effective potential from any diagram can arise only if number of
$D^2$-factors is more by one than the number of $\bar{D}^2$-factors.

The only Green function in the theory is a propagator
$<\phi\bar{\phi}>$. Therefore total number of quantum chiral superfields
$\phi$ corresponding to all vertices
must be equal to that one of antichiral ones $\bar{\phi}$.

Now we turn directly to studying of supergraphs given in Figs. 1a, 1b.
First let us consider diagrams of the form Fig.1a. The only vertex
of fourth order in quantum superfields has the form
$\phi^l\bar{\phi}^{4-l}$ where $l$ is a some integer taking values from
0 to 4.
Each vertex proportional to $W^{''}$ corresponds to two
$\phi$ and each one proportional to $K_{\bar{\Phi}\bar{\Phi}}$
corresponds to two $\bar{\phi}$. Then, vertex proportional to
$K_{\Phi\bar{\Phi}}-1$ corresponds to one $\phi$ and one $\bar{\phi}$.
Therefore total number of chiral quantum superfields $\phi$ in vertices leading
to such a diagram is equal to
$l+2 n_{W^{''}}+n_{K_{\Phi\bar{\Phi}}}$ and that one of antichiral quantum
superfields
$\bar{\phi}$ -- to $4-l+2 n_{K_{\bar{\Phi}\bar{\Phi}}}+n_{K_{\Phi\bar{\Phi}}}$,
where $n_{W^{''}}$, $n_{K_{\Phi\bar{\Phi}}}$ and $n_{K_{\bar{\Phi}\bar{\Phi}}}$  --
are numbers of vertices proportional to
$W^{''}$, $K_{\Phi\bar{\Phi}}-1$ and $K_{\bar{\Phi}\bar{\Phi}}$ respectively.
Since numbers of quantum superfields $\phi$ and $\bar{\phi}$
corresponding to all vertices must be equal we conclude that
$l+2 n_{W^{''}}=4-l+2 n_{K_{\bar{\Phi}\bar{\Phi}}}$.
Hence $n_{K_{\bar{\Phi}\bar{\Phi}}}=n_{W^{''}}+l-2$.
Then, let us find number of factors $D^2$ and $\bar{D}^2$ in such a
diagram.  The vertex proportional to $\phi^l\bar{\phi}^{4-l}$
corresponds to $l$ factors $\bar{D}^2$ and $4-l$ factors $D^2$.
However, if $l=0$ or $l=4$, number of factors $D^2$ (respectively
$\bar{D}^2$) must be reduced by one because transformation of the
expression corresponding to such a vertex to the form of integration
over total superspace by the rule
$\int d^2\theta(-\frac{\bar{D}^2}{4})=\int d^4\theta$
(or $\int d^2\bar{\theta}(-\frac{D^2}{4})=\int d^4\theta$)
requires one $\bar{D}^2$- (respectively $D^2$-)factor. Therefore
the vertex of fourth order in quantum superfields corresponds to
$l-N_{V}$ $\bar{D}^2$-factors and $4-l-N_{\bar{V}}$
$D^2$-factors where $N_{V}$ and $N_{\bar{V}}$ are numbers of purely chiral and
antichiral vertices of third and higher orders in quantum superfields
($N_{V}$ and $N_{\bar{V}}$ can be equal to
0 or 1 for such a supergraph).
Besides, each vertex proportional to $K_{\bar{\Phi}\bar{\Phi}}$ corresponds to
two $D^2$-factors, and that one proportional to $W^{''}$ -- to one
$\bar{D}^2$-factor.
As a result, total number of $D^2$-factors is equal to
$2 n_{K_{\bar{\Phi}\bar{\Phi}}}+4-l-N_{\bar{V}}
+n_{K_{\Phi\bar{\Phi}}}$, and that one of $\bar{D^2}$-factors -- to
$n_{W^{''}}+l-N_{V}+n_{K_{\Phi\bar{\Phi}}}$.
We again note that the
contribution to holomorphic effective potential can arise if and only if
total number of $D^2$-factors is more by one than number of
$\bar{D}^2$-factors.
Then, $n_{K_{\bar{\Phi}\bar{\Phi}}}=n_{W^{''}}+l-2$ (see above).
Therefore necessary correlation between numbers of
$D^2$- and $\bar{D}^2$-factors is satisfied if
$n_{W^{''}}+N_{V}-N_{\bar{V}}=1$. However, we proved that
non-trivial holomorphic effective potential can arise if $n_{W^{''}}$
is equal to 3, 4 or 5. And $N_{V}$ and $N_{\bar{V}}$ can be equal to 0
or 1 for supergraphs of such a type.  Therefore relation
$n_{W^{''}}-N_{V}+N_{\bar{V}}=1$ is not satisfied and diagrams of the type
given in Fig.1a do not contribute to holomorphic effective potential.

Now we turn to studying of diagrams of the type written in Fig.1b.
These supergraphs contain two vertices which have the form
$\phi^{l_1}\bar{\phi}^{3-l_1}$ and $\phi^{l_2}\bar{\phi}^{3-l_2}$ respectively.
The $l_1,l_2$ are integer, they can take values 0,1,2,3.
In analogy with the previous case, it can be shown that the total number of
chiral quantum superfields $\phi$ in vertices leading such a diagram is equal to
$l_1+l_2+2 n_{W^{''}}+n_{K_{\Phi\bar{\Phi}}}$, and total number of
antichiral quantum superfields $\bar{\phi}$ -- to
$4-l+2 n_{K_{\bar{\Phi}\bar{\Phi}}}+n_{K_{\Phi\bar{\Phi}}}$. Since these two
numbers must be equal (see above) we conclude that
$n_{K_{\bar{\Phi}\bar{\Phi}}}=n_{W^{''}}+l_1+l_2-3$.
Further, number of $D^2$-factors is equal to
$2 n_{K_{\bar{\Phi}\bar{\Phi}}}+6-l_1-l_2-N_{\bar{V}}
+n_{K_{\Phi\bar{\Phi}}}$, and the number of $\bar{D^2}$-factors -- to
$n_{W^{''}}+l_1+l_2-N_{V}+n_{K_{\Phi\bar{\Phi}}}$. As usual, any diagram can
contribute to holomorphic effective potential if number of
$D^2$-factors is more by one than number of $\bar{D}^2$-factors.
Since $n_{K_{\bar{\Phi}\bar{\Phi}}}=n_{W^{''}}+l_1+l_2-3$ non-trivial
contribution to holomorphic effective potential can arise if
$n_{W^{''}}+N_{V}-N_{\bar{V}}=1$. This condition is satisfied only if
$n_{W^{''}}=3$, $N_{V}=0$, $N_{\bar{V}}=2$.
Therefore both vertices of third order in quantum superfields have the form
$\bar{W}^{'''}\bar{\phi}^3$, and $n_{K_{\bar{\Phi}\bar{\Phi}}}=0$.
We note that there is no any restrictions for the number of vertices
proportional to $K_{\Phi\bar{\Phi}}-1$.
The possible supergraph satisfying these conditions includes three
external vertices $W^{''}$ and two antichiral vertices of the form
$\bar{W}^{'''}\bar{\phi}^3$.  There is the only diagram of such a form
(see Fig.3).


Now let us turn straightforward calculation of two-loop
correction
to holomorphic effective potential. We have shown that
supergraphs contributing to this potential
can contain only vertices proportional to $W^{''}$, $\bar{W}^{'''}$,
$K_{\Phi\bar{\Phi}}-1$. It means it is sufficient to consider the
theory with the action
\begin{eqnarray}
\label{qchr}
S=\int d^8 z K_{\Phi\bar{\Phi}}\phi\bar{\phi}+
\int d^6 z \frac{1}{2}W^{''}\phi^2+\int d^6 \bar{z}\frac{1}{3!}
\bar{W}^{'''}\bar{\phi}^3
\end{eqnarray}
We find the propagator of quantum superfield $\phi$ corresponding to
quadratic action
$S_0=\int d^8 z K_{\Phi\bar{\Phi}}\phi\bar{\phi}$. It can be represented
in the form of expansion in vertices $K_{\Phi\bar{\Phi}}-1$ which has the form
(Fig.2)

\vspace*{2mm}

\Lengthunit=1.4cm
\GRAPH(hsize=2){
\Linewidth{1.2pt}
\lin(1,0)
\ind(1,3){D^2}\ind(7,3){\bar{D}^2}\ind(2,0){|}\ind(7,0){|}
\ind(12,0){=}
\hspace{.3cm}
\mov(1.2,0){\GRAPH(hsize=5){
\Linewidth{0.3pt}
\lin(1,0)
\ind(1,3){D^2}\ind(7,3){\bar{D}^2}\ind(2,0){|}\ind(7,0){|}
\ind(15,0){+}\mov(2,0){\lin(2,0)}\mov(3,0){\dashdotlin(0,1)}
\ind(27,13){K_{\Phi\bar{\Phi}}-1}
\ind(21,3){D^2}\ind(26,3){\bar{D}^2}\ind(22,0){|}\ind(27,0){|}
\ind(29,3){D^2}\ind(37,3){\bar{D}^2}\ind(30,0){|}\ind(37,0){|}
\ind(45,0){+}
\mov(4.8,0){\lin(3,0)}\mov(6,0){\dashdotlin(0,1)}\mov(7,0){\dashdotlin(0,1)}
\ind(55,13){K_{\Phi\bar{\Phi}}-1}\ind(72,13){K_{\Phi\bar{\Phi}}-1}
\ind(49,3){D^2}\ind(54,3){\bar{D}^2}\ind(50,0){|}\ind(55,0){|}
\ind(59,3){D^2}\ind(64,3){\bar{D}^2}\ind(58,0){|}\ind(64,0){|}
\ind(69,3){D^2}\ind(75,3){\bar{D}^2}\ind(67,0){|}\ind(73,0){|}
\ind(76,0){+}\ind(80,0){\ldots}\ind(35,-5){Fig.2}
}}}

\vspace*{1mm}

\noindent Bold line in this supergraph denotes the propagator
corresponding to quadratic action (\ref{qchr}), and thin one is
standard superpropagator (\ref{pr1}). Dashed-and-dotted line means the
external superfield $K_{\Phi\bar{\Phi}}-1$.

After summarization of the chain written in Fig.2 the propagator depending on
background superfields looks like
\begin{equation}
\label{propc}
<\phi\bar{\phi}>=-\bar{D}^2_1 D^2_2
\frac{\delta^8(z_1-z_2)}{16 K_{\Phi\bar{\Phi}}(z_1)\Box}
\end{equation}
We note that the superfield $K_{\Phi\bar{\Phi}}$ is not constant here.

We have already pointed that there is the only supergraph
contributing to holomorphic effective
potential in two-loop approximation. This diagram is written in Fig.3.
Double external lines here denote background superfields $W^{''}$.


\hspace{4.5cm}
\Lengthunit=1.5cm
\Linewidth{1.2pt}
\GRAPH(hsize=3){\ind(0,-16){Fig.3}
\mov(.5,0){\Circle(2)\mov(-1,0){\lin(2,0)}
\ind(-2,10){|}
\ind(-2,-3){\bar{D}^2}\ind(-2,0){|}\ind(-2,-10){|}\ind(-2,-13){\bar{D}^2}
\ind(-2,7){\bar{D}^2}
\ind(-9,2){D^2} \ind(-10,-2){D^2} \ind(8,2){D^2} \ind(8,-2){D^2}
\ind(-18,2){-} \ind(-18,-2){-} \ind(0,2){-} \ind(0,-2){-}
\Linewidth{0.3pt}
\mov(-1,1){\lin(-.7,.7)}\mov(-1.1,1){\lin(-.7,.7)}
\mov(-1,-1){\lin(.7,-.7)}\mov(-1.1,-1){\lin(.7,-.7)}
\mov(-1,0){\lin(-.7,.7)}\mov(-1.1,0){\lin(-.7,.7)}}
}

\vspace*{2mm}

\noindent Contribution of supergraph given in Fig.3 looks like
\begin{eqnarray}
\label{I1}
I&=&\int \frac{d^4p_1 d^4p_2}{{(2\pi)}^8}\frac{d^4k d^4l}{{(2\pi)}^8}
\int d^4\theta_1 d^4\theta_2 d^4\theta_3 d^4\theta_4 d^4\theta_5
\frac{W^{''}}{K^2_{\Phi\bar{\Phi}}}(-p_1,\theta_3)
\frac{W^{''}}{K^2_{\Phi\bar{\Phi}}}(-p_2,\theta_4)\times\nonumber\\&\times&
\frac{W^{''}}{K^2_{\Phi\bar{\Phi}}}(p_1+p_2,\theta_5)
{\bar{W}}^{'''2}
\frac{1}{k^2 l^2 {(k+p_1)}^2{(l+p_2)}^2{(l+k)}^2{(l+k+p_1+p_2)}^2}
\times\nonumber\\&\times&
\delta_{13}\frac{\bar{D}^2_3}{4}\delta_{32}
\frac{D^2_1 \bar{D}^2_4}{16}\delta_{14}\delta_{42}
\frac{D^2_1 \bar{D}^2_5}{16}\delta_{15}\delta_{52}
\end{eqnarray}
After $D$-algebra transformation this expression can be written as
\begin{eqnarray}
\label{app}
I&=&\int \frac{d^4p_1 d^4p_2}{{(2\pi)}^8}\frac{d^4k d^4l}{{(2\pi)}^8}
\int d^2\theta
{\bar{W}}^{'''2}
\frac{W^{''}}{K^2_{\Phi\bar{\Phi}}}(-p_1,\theta)
\frac{W^{''}}{K^2_{\Phi\bar{\Phi}}}(-p_2,\theta)
\frac{W^{''}}{K^2_{\Phi\bar{\Phi}}}(p_1+p_2,\theta)
\times\nonumber\\&\times&
\frac{k^2 p_1^2+ l^2 p_2^2 +2 (k l)(p_1 p_2)}
{k^2 l^2 {(k+p_1)}^2{(l+p_2)}^2{(l+k)}^2{(l+k+p_1+p_2)}^2}
\end{eqnarray}

As we know the effective potential is the
effective lagrangian for superfields slowly varying in space-time.
Let us study behaviour of the expression (\ref{app}) in this case.
The contribution (\ref{app}) can be expressed as
\begin{eqnarray}
\label{cont}
I&=&\int d^2\theta \int \frac{d^4p_1 d^4p_2}{{(2\pi)}^8}
{\bar{W}}^{'''2}
\frac{W^{''}}{K^2_{\Phi\bar{\Phi}}}(-p_1,\theta)
\frac{W^{''}}{K^2_{\Phi\bar{\Phi}}}(-p_2,\theta)
\frac{W^{''}}{K^2_{\Phi\bar{\Phi}}}(p_1+p_2,\theta)
S(p_1, p_2)
\end{eqnarray}
Here $p_1, p_2$ are external momenta. The expression $S(p_1, p_2)$ here
is equal to
$$ \int\frac{d^4 k d^4 l}{{(2\pi)}^8}\frac{k^2 p_1^2+ l^2
p_2^2 +2 (kl)(p_1 p_2)} {k^2 l^2
{(k+p_1)}^2{(l+p_2)}^2{(l+k)}^2{(l+k+p_1+p_2)}^2} $$
After Fourier transform eq. (\ref{cont}) has the form
\begin{eqnarray}
\label{cont1}
I&=&\int d^2\theta \int d^4 x_1 d^4x_2 d^4 x_3
\int\frac{d^4p_1 d^4p_2}{{(2\pi)}^8}
{\bar{W}}^{'''2}
\frac{W^{''}}{K^2_{\Phi\bar{\Phi}}}(x_1,\theta)
\frac{W^{''}}{K^2_{\Phi\bar{\Phi}}}(x_2,\theta)\times\nonumber\\&\times&
\frac{W^{''}}{K^2_{\Phi\bar{\Phi}}}(x_3,\theta)
\exp[i(-p_1 x_1- p_2 x_2+(p_1+p_2)x_3)]
S(p_1, p_2)
\end{eqnarray}
Since superfields in the case under consideration are slowly varying in
space-time we can put
$
\frac{W^{''}}{K^2_{\Phi\bar{\Phi}}}(x_1,\theta)
\frac{W^{''}}{K^2_{\Phi\bar{\Phi}}}(x_2,\theta)
\frac{W^{''}}{K^2_{\Phi\bar{\Phi}}}(x_3,\theta)
\simeq {\Big\{\frac{W''(x_1,\theta)}{K_{\Phi\bar{\Phi}}^2(x_1,\theta)}\Big\}}^3
$.
As a result one gets
\begin{eqnarray}
I&=&\int d^2\theta \int d^4 x_1 d^4 x_2 d^4 x_3
\int\frac{d^4p_1 d^4p_2}{{(2\pi)}^8}{\bar{W}}^{'''2}
{\Big\{\frac{W''(x_1,\theta)}{K_{\Phi\bar{\Phi}}^2(x_1,\theta)}\Big\}}^3
\times\nonumber\\&\times&
\exp[i(-p_1 x_1- p_2 x_2+(p_1+p_2)x_3)]
S(p_1, p_2)
\end{eqnarray}
Integration over $d^4 x_2 d^4 x_3$ leads to delta-functions
$\delta(p_2)\delta(p_1+p_2)$. Hence the eq. (\ref{cont1}) takes the form
\begin{equation}
I=\int d^2\theta \int d^4 x_1 {\bar{W}}^{'''2}
{\Big\{\frac{W''(x_1,\theta)}{K_{\Phi\bar{\Phi}}^2(x_1,\theta)}\Big\}}^3
S(p_1,p_2)|_{p_1,p_2=0}
\end{equation}
Therefore final result for two-loop correction to holomorphic effective
potential looks like
\begin{equation}
\label{l2c}
W^{(2)}=
\frac{6}{{(16\pi^2)}^2}\zeta(3) \bar{W}^{'''2}
{\Big\{\frac{W''(z)}{K^2_{\Phi\bar{\Phi}}(z)}\Big\}}^3
\end{equation}
One reminds that $\bar{W}^{'''}=\bar{W}^{'''}(\bar{\Phi})|_{\bar{\Phi}=0}$ and
$K_{\Phi\bar{\Phi}}(z)=\frac{\partial ^2 K(\bar{\Phi},\Phi)}
{\partial\Phi\partial\bar{\Phi}}
|_{\bar{\Phi}=0}$ here.
We also took into account that
$$
\int \frac{d^4k d^4l}{{(2\pi)}^8}
\frac{k^2 p_1^2+ l^2 p_2^2 +2 (k_1 k_2)(p_1 p_2)}
{k^2 l^2 {(k+p_1)}^2{(l+p_2)}^2{(l+k)}^2{(l+k+p_1+p_2)}^2}|_{p_1=p_2=0}
=\frac{6}{{(4\pi)}^4}\zeta(3)
$$
We see that the correction (\ref{l2c}) is finite and does not require
renormalization.  It is evident that eq. (\ref{l2c}) reproduces
the known results for Wess-Zumino model
at $W=-\frac{\lambda}{3!}\Phi^3$, $K=\bar{\Phi}\Phi$
\cite{West2,West3,Buch3,Buch4}.

We would like to emphasize that the result (\ref{l2c}) defining holomorphic
effective potential corresponds to only massless theory. Therefore, the known
statement \cite{Se} about absence of
holomorphic corrections in Wess-Zumino model which has been obtained for massive
theory, cannot be applied in the case under consideration.

To conclude, we have solved the problem of calculating leading holomorphic
correction to superfield effective action in general chiral superfield model
(1) with arbitrary potentials $K(\bar{\Phi},\Phi)$ and $W(\Phi)$.
The result has the universal form (\ref{l2c}) and it is finite independently
if the functions $K(\bar{\Phi},\Phi)$, $W(\Phi)$ correspond to
renormalizable theory or no.

{\bf Acknowledgements.} Authors are grateful to M. Cvetic and S.M. Kuzenko
for discussions. The work was carried out under
partial support of INTAS, project INTAS--96--0308; RFBR -- DFG, project No.
96-02-00180; RFBR, project No. 99-02-16617; grant center of
St. Peterburg University, project No. 97--6.2--34.

\end{document}